\newif\ifdraft
\draftfalse

\newif\ifrev
\revfalse

\newif\ifflotsam
\flotsamfalse

\documentclass[12pt]{article}
\usepackage{graphicx,amsmath,bm}
\usepackage[numbers,square,sort&compress]{natbib}
\usepackage[margin=2.5cm]{geometry}

\usepackage{url}
\usepackage{breakurl}

\usepackage[breaklinks]{hyperref}
\usepackage[usenames,dvipsnames]{xcolor}
\usepackage{ifthen}
\hypersetup{
    colorlinks,
    linkcolor={red!50!black},
    citecolor={blue!50!black},
    urlcolor={blue!80!black}
}
\usepackage[normalem]{ulem}
\ifdraft 
    \newcommand{\NBB}[1]{{\color{MidnightBlue}[NBB: #1]}}
    
    \newcommand{\HK}[1]{{\color{magenta}[HK: #1]}}
    
    \newcommand{\crossout}[1]{{\color{red}\sout{#1}}}
\else   
    \newcommand{\NBB}[1]{}
    
    \newcommand{\HK}[1]{}
    
    \newcommand{\crossout}[1]{}   
\fi 

\newcommand{\fig}[2]{
    \begin{figure}
        \begin{center}
            \includegraphics[width=0.49\textwidth]{fig/#1}
        \end{center}
    \caption{#2 \label{fig/#1}}
    \end{figure}
    }

\newcommand{\fullfig}[2]{
    \begin{figure*}
        \begin{center}
            \includegraphics[width=0.8\textwidth]{fig/#1}
        \end{center}
    \caption{#2 \label{fig/#1}}
    \end{figure*}
    }

\newcommand{\figref}[1]{Figure\,\ref{fig/#1}}

\newcommand{\fulltab}[5]{ 
    \begin{table*}[ht]     
        \centering
        \caption{
            #2 \label{tab/#1}
        }    
        \begin{tabular}{#3} 
        \hline 
            #4 \\
        \hline
         #5 
        \end{tabular}
    \end{table*}
}

\newcommand{\tabref}[1]{Table\,\ref{tab/#1}}

\newcommand{\eq}[2]{
    \begin{equation}
        #2 \label{#1}
    \end{equation}
}

\renewcommand{\~}[1]{\tilde{#1}}

\newcommand{\al}{\ensuremath{\alpha}}

\newcommand{\be}{\ensuremath{\beta}}

\newcommand{\ga}{\ensuremath{\gamma}}

\newcommand{\de}{\ensuremath{\delta}}
\newcommand{\DE}{\ensuremath{\Delta}}
\newcommand{\ep}{\ensuremath{\epsilon}}

\newcommand{\et}{\ensuremath{\eta}}

\newcommand{\la}{\ensuremath{\lambda}}

\newcommand{\rh}{\ensuremath{\rho}}

\newcommand{\si}{\ensuremath{\sigma}}

\newcommand{\ta}{\ensuremath{\tau}}

\newcommand{\PH}{\ensuremath{\Phi}}

\newcommand{\pRe}{\textit{Re}}
\newcommand{\dt}{\ensuremath{\Delta t}}

\newcommand{\vu}{\ensuremath{\bm{u}}}

\renewcommand{\d}{\partial}

\begin{document}

\title{
    Scale-dependent Error Growth in Navier--Stokes Simulations
}

\author{
    Nazmi Burak Budanur and Holger Kantz \\
    \small{Max Planck Institute for the Physics of Complex Systems (MPIPKS)}\\
    \small{Nöthnitzer Straße 38, 01187 Dresden, Germany}
}
\date{\today}

\newcommand{\alnum}{0.18}
\newcommand{\rhnum}{0.32}

\maketitle	

\begin{abstract}
    We estimate the maximal Lyapunov exponent at different resolutions and
    Reynolds numbers in large eddy (LES) and direct numerical simulations (DNS)
    of sinusoidally-driven Navier--Stokes equations in three dimensions.
    Independent of the Reynolds number when nondimensionalized by Kolmogorov
    units, the LES Lyapunov exponent diverges as an inverse power of the
    effective grid spacing showing that the fine scale structures exhibit much
    faster error growth rates than the larger ones.
    Effectively, i.e., ignoring
    the cut-off of this phenomenon at the Kolmogorov scale, this behavior
    introduces an upper bound to the prediction horizon that can be achieved by
    improving the precision of initial conditions through refining of the
    measurement grid.
\end{abstract}

\section{Introduction}

A defining feature of deterministic chaos is that infinitesimal errors in the
initial conditions grow, on average, exponentially fast in time, the rate being
called the maximal Lyapunov exponent $\la$ of the system. This is a challenge to
predictions and often attributed to the difficulties in weather forecasting
since the initial condition which one plugs into some model equations will
represent the reality only up to some inaccuracy and even if the model were
perfect, the initial error $\ep_0$ would grow approximately as $\ep(t)\approx
\ep_0 e^{\la t}$. At latest when the distance between this forecast trajectory
and the unknown ``true'' trajectory of the system reaches the order of magnitude
of the attractor size (or, more pragmatically, the standard deviation of the
quantity to be forecast), this forecast has lost its usefulness, and the
prediction horizon is reached. This exponential error growth is already quite
unfavorable if one wants to extend the prediction horizon: for a linear
extension $\de t$ of the latter, the initial inaccuracy has to be reduced
exponentially by a factor of $e^{-\la \de t}$. 

Since the 1950s, several authors
\cite{thompson1957uncertainty,lorenz1969predictability,robinson1967some,
robinson1971predictability,leith1972predictability,aurell1996growth,
rotunno2008generalization,boffetta2017chaos} 
have argued that the practical situation in forecasting hydrodynamic systems is likely 
to be much worse when a range of spatial and temporal scales are at present as in the
case of turbulent flows \cite{richardson2007weather,kolmogorov1991local}. 
The basic idea is that for a multi-scale hydrodynamic system, 
refining the initial
conditions would necessarily translate to the inclusion of small-scale motions
into the model, which in turn have shorter characteristic temporal scales,
such as the eddy turnover time \cite{pope2000turbulent}, thus resulting in 
faster-growing errors. While 
\cite{thompson1957uncertainty,lorenz1969predictability,robinson1967some,
robinson1971predictability,leith1972predictability,aurell1996growth,
rotunno2008generalization,boffetta2017chaos} 
differ in many aspects such as the systems and models that they consider, 
their common conclusion is that the predictability in hydrodynamic systems 
is much more strongly limited than by the simple exponential divergence of 
trajectories as a result of the described phenomenon, which hereafter we refer 
to as \emph{scale-dependent error growth}. 

Evidence of scale-dependent error growth has also been reported in the models 
of atmosphere: In \cite{harlim2005convex}, fast-growth of errors at small-scales 
was found in the  
Global Forecast System of the National Centers for Environmental Prediction, 
and more recently in \cite{bednar2021prediction} for ensemble 
forecasts of the European Centre for Medium-Range Weather Forecasts. 
In \cite{brisch2019power}, it was argued that the results of
\cite{harlim2005convex} were compatible with the power-law divergence 
of the finite-size Lyapunov exponent in vanishing amplitude of the 
initial error. Altogether, these results point to a significant role of 
scale-dependent error growth for weather forecasts.

The above-cited papers on scale-dependent error growth consider forecast
scenarios in which one is interested in predicting large-scale motions in
the system which are assumed to be measured with high accuracy and asks ``How
long will it take for the uncertainties in unresolved length scales to
contaminate those of interest?'' This question differs fundamentally from
that of the standard exponential divergence of infinitesimal errors as it
is concerned with finite-size deviations. Motivated by this, 
Aurell et al. \cite{aurell1996growth} introduced the 
concept of finite size Lyapunov exponents and applied it to a shell model 
turbulence cascade. 
By computing the error doubling times for perturbations of a finite size 
$E$ and averaging over the system's attractor, they concluded that the rate 
$\la_F (E)$ of error growth can be well described by a power law 
$\la_F (E) \propto E^{- \be}$ with $\be > 0$. 
Later, Boffetta and Musacchio \cite{boffetta2017chaos} presented numerical 
evidence for this kind of error growth behavior in the direct numerical 
simulations (DNS) of homogeneous isotropic turbulence. 

In this paper, we present simulation results which demonstrate that the
scale-dependent error growth is already relevant for the growth rate of 
infinitesimal perturbations in hydrodynamic models with varying resolutions.
In order to understand how inclusion of smaller scales affects the
rate of error growth, we perform large eddy simulations (LES) at different
resolutions, wherein the motions at length scales smaller than the grid spacing
are explicitly filtered out. In these, we estimate the maximal Lyapunov
exponents and compare to those we find in the fully-resolved direct numerical
simulations (DNS). Our results confirm that increasing the spatial resolution of
LES indeed results in faster-growing errors. Moreover, when normalized by the
Kolmogorov length and time scales, the rates of error growth at different
Reynolds numbers and resolutions collapse on a single curve that can be 
fit by a power law.
Finally, we consider a forecast scenario where the error on the initial 
state's measurement is proportional to the grid spacing of LES and show 
that scale-dependent error growth yields an upper limit for the maximal prediction 
horizon that can be achieved through LES. 

\section{Computational setup}

Let us denote the three-dimensional cartesian coordinates by
$(x_1, x_2, x_3)$ and the partial derivatives with respect to 
time $t$ and $x_i$ as $\d_t$ and $\d_i$, respectively.  
Adopting also the summation convention over the repeated 
indices, the filtered Navier--Stokes equations read
\eq{navier-stokes}{
	\d_t u_j = 
        - u_i  \d_i u_j - \d_j p 
        + 2 \d_i (\pRe^{-1} + \nu_T) S_{ij} + f_j \, , 
}
where $u_j$ are the velocity field components
subjected to the incompressibility $\d_j u_j = 0$  and 
periodic boundary conditions $u_j|_{x_i + L_i} = u_j|_{x_i}$, 
\pRe\ is the Reynolds number, 
$S_{ij} = (\d_i u_j + \d_j u_i) / 2$ is the rate-of-strain tensor, 
$f_j = (4 \pRe)^{-1} \sin (2 \pi x_2 / L_2) \de_{1j}$ is a sinusoidal body force, 
and $\nu_T$ is the so-called ``eddy viscosity'' that aims to account for 
the transfer of energy to the length scales below a resolution. 
We implement the Smagorinsky model \cite{smagorinsky1963general}
\eq{eddy_viscosity}{
    \nu_T = (C_S \DE)^2 \sqrt{S_{ij} S_{ij}} \,,
}
where $C_S$ is the Smagorinsky constant and $\DE$ is the length scale below 
which the motion is not resolved. Note that the full Navier--Stokes equations 
corresponding to the three-dimensional Kolmogorov flow \cite{shebalin1997kolmogorov}
are recovered in \eqref{navier-stokes} if $\nu_T = 0$.

We simulate \eqref{navier-stokes} using \texttt{dnsbox} 
\cite{yalniz2021coarse,yalniz2021dnsbox} which discretizes the velocity 
field $u_i$ using Fourier series in all three directions, compute the nonlinear 
terms in \eqref{navier-stokes} pseudospectrally \cite{canuto2007spectral} and 
time-step the equations using a semi-implicit second-order predictor-corrector 
scheme. All of our simulations are carried out in a cubic domain with the edge 
length $L_i = 4$, yielding the laminar solution of \eqref{navier-stokes} when 
$\nu_T=0$ as 
\(
    u_j = \sin (\pi x_2 / 2) \de_{1j} 
\). 
The laminar solution with a unit peak amplitude at $x_2 = 1$ sets 
our length and velocity scales, thus the time scale as their ratio. 

\tabref{domains} shows a summary of our simulation domains. For the DNS domains,
we determine our resolutions $n^{DNS}$ such that the grid spacing 
$L_i / n^{DNS}$ is on the order of the Kolmogorov length scale $\et = (\pRe^{3}
\varepsilon)^{-1/4}$, where $\varepsilon$ is the rate of turbulence dissipation.
We estimate this as $\varepsilon = \pRe^{-1} \langle \d_j \~{u}_i \, \d_j
\~{u}_i \rangle$ where $\~{u}_i$ are the fluctuating velocity components with
means along the homogeneous directions $x_1$ and $x_3$ subtracted and the angle
brackets indicate the average over  simulation domain. 
Besides the \pRe\ based on the laminar solution, in \tabref{domains} we 
also report $\pRe_{rms} = \langle \tilde{u} \rangle_{rms} L_i \pRe $ and 
$\pRe_{T} = \langle \tilde{u} \rangle_{rms} l_{T} \pRe$ where 
$\langle \tilde{u} \rangle_{rms}$ is the root mean square of the fluctuating 
velocity and 
$l_{T} = \sqrt{15 / (\varepsilon \pRe)} \langle \tilde{u} \rangle_{rms} \pRe$ 
is the Taylor microscale \cite{pope2000turbulent}.
Our final estimates of $\et$, $\pRe_{rms}$ and $\pRe_T$
come from an average of it over $10$ fluid states in each DNS domain. 
We started our simulations with a variable time-step that keeps the 
Courant–Friedrichs–Lewy number approximately equal to $0.25$, once a state 
on the attractor is reached, we fix our time step to $\dt = 0.04$ in DNS
and $\dt = 0.02$ in LES domains.

\fulltab{domains}{
    Properties of the simulation domains. 
    In all LES simulations the Smagorinsky constant was set to $C_S = 0.1$.
    $\min n^{LES}$ and  $\max n^{LES}$ corresponds to the smallest and the 
    largest number of grid points used in LES at the corresponding $\pRe$. 
    $k_{max} = 2 \pi (n^{DNS} / 2 + 1) / L_i$ is the largest resolved wave 
    number in each direction. 
}{
    c c c c c c c
}{
    $\pRe$ & $\pRe_{rms}$ & $\pRe_{T}$ &  $\min n^{LES}$ & $\max n^{LES}$ & $n^{DNS}$ & $k_{\max} \et$  
}{
    $10{,}000$ & $460.41     \,\pm\, 45.80$ & $30.07  \,\pm\, 4.41$ & $24$ & $80$ & $216$ & $4.07$ \\
    $20{,}000$ & $696.53     \,\pm\, 57.34$ & $38.45  \,\pm\, 4.16$ & $24$ & $96$ & $240$ & $3.39$ \\
    $30{,}000$ & $866.35     \,\pm\, 78.90$ & $41.60  \,\pm\, 3.41$ & $24$ & $112$ & $288$ & $3.41$ \\
    $40{,}000$ & $1{,}006.55 \,\pm\, 72.88$ & $45.14  \,\pm\, 4.15$ & $24$ & $128$ & $320$ & $3.39$ \\
    $80{,}000$ & $1{,}412.58 \,\pm\, 94.68$ & $54.19  \,\pm\, 6.30$ & $24$ & $144$ & $432$ & $3.58$ 
}

Let  $\vu (t)$ be the fluid state at time $t$,    
$\PH^{t}$ denote the time-$t$ map induced by the dynamics transforming 
$\vu(t)$ as  
\( 
    \vu (t) = \PH^{t} (\vu (0))
\),  and $\| . \|$ indicate the $L_2$ norm 
$\|\vu\| = (L_1 L_2 L_3)^{-1}$ $\int \int \int u_i u_i d^3 x$.
We estimate the maximal Lyapunov exponents via Benettin algorithm 
\cite{benettin1980lyapunov} through the following steps. 
Starting from an initial state $\vu (0)$ on the attractor, we 
add a random perturbation $\de \vu (0)$ with 
$\| \de \vu (0) \| = \si \| \vu (0) \|$, 
where $\si$ is a small positive real number, to this state and simulate  
$\vu(t)$ and $\vu (t) + \de \vu (t)$ in 
parallel. At each $\ta$, we rescale $\de \vu (n \ta)$ such that its 
norm is set as 
$\| \de \vu (n \ta) \| = \si \| \vu (n \ta) \|$ where 
$n = 1, 2, \ldots$ After some transient time $t_{trans} = n_{t} \ta$, we 
estimate the maximal Lyapunov exponent as
\eq{max_lyap_exp}{
    \lambda \approx \frac{1}{N \ta} \sum_{n = n_{t}}^{n_{t} + N} 
    \ln \frac{
        \| \PH^{\ta} (\vu (n \ta) + \de \vu (n \ta)) 
         - \PH^{\ta} (\vu (n \ta)) \|}{
              \|              \de \vu (n \ta)  \|
        }\,.
}
For our numerical results to follow, we used 
$\si = 10^{-4}$, $t_{trans} = 400$, and 
$\ta = 20$ in DNS and 
$\si = 10^{-4}$, $t_{trans} = 400$, 
and $\ta = 5$ in LES. 
We confirmed that our results are insensitive to these parameter choices by 
partially reproducing them using $\si$ one order of magnitude up and down; 
and $\DE t$ at half and double of its value. 

\section{Results}
\label{sec/results}

\fig{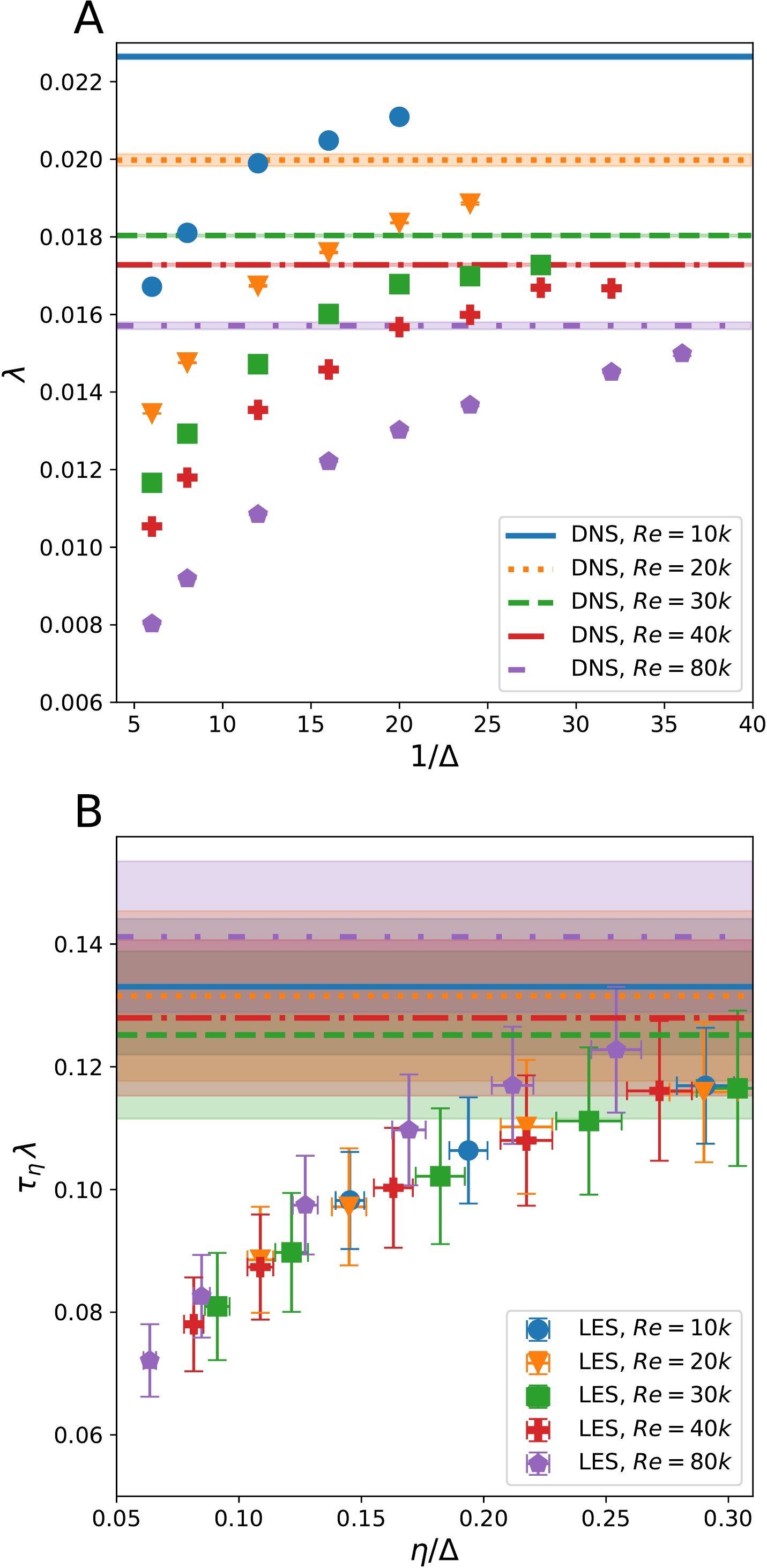}{
    (color online) 
    \textbf{A.} Maximal Lyapunov exponent estimated in LES at different
    resolutions.
    Fully-resolved DNS Lyapunov exponents are indicated by horizontal lines.
    \textbf{B.} Same as A, scaled by
    Kolmogorov length and time scales. Legends in A and B apply to both panels.
}

\figref{lyap_les_dns}A shows the numerical estimates of the largest Lyapunov 
exponent obtained in DNS at five different $\pRe$ 
as straight lines, and the resolution-dependent exponents obtained in 
LES at the same set of $\pRe$s against $1/\DE$. Shown values are obtained by 
averaging over the estimates of the final $1000$ time units in each simulation 
and the widths of error bars and shaded regions in 
\figref{lyap_les_dns} correspond to two standard deviations. We observe that 
as the LES resolution is increased, so does the Lyapunov exponent 
with a clear trend towards the corresponding exponent from DNS.
Note that the decrease of Lyapunov exponents as 
$\pRe$ is increased is due to our nondimensionalization of the system 
through the sinusoidal laminar solution of the Kolmogorov flow. 
When we rescale the Lyapunov exponents, which are measured in inverse 
time units, by the Kolmogorov time 
$\ta_\et = (\pRe\, \varepsilon)^{-1/2}$, and plot them against the LES 
grid spacing $\Delta$ normalized by the 
 the Kolmogorov length $\eta$, 
we obtain the data collapse shown in \figref{lyap_les_dns}B. Hence, 
the Reynolds-number dependence is fully accounted for by the Reynolds-number 
dependence of the Kolmororov time and length scale. 
The larger uncertainties in \figref{lyap_les_dns}B with respect to those 
in \figref{lyap_les_dns}A are due to the standard deviations of our estimates 
of $\et$ and $\ta$, which we compute as an ensemble average over turbulent DNS 
states. 

\fullfig{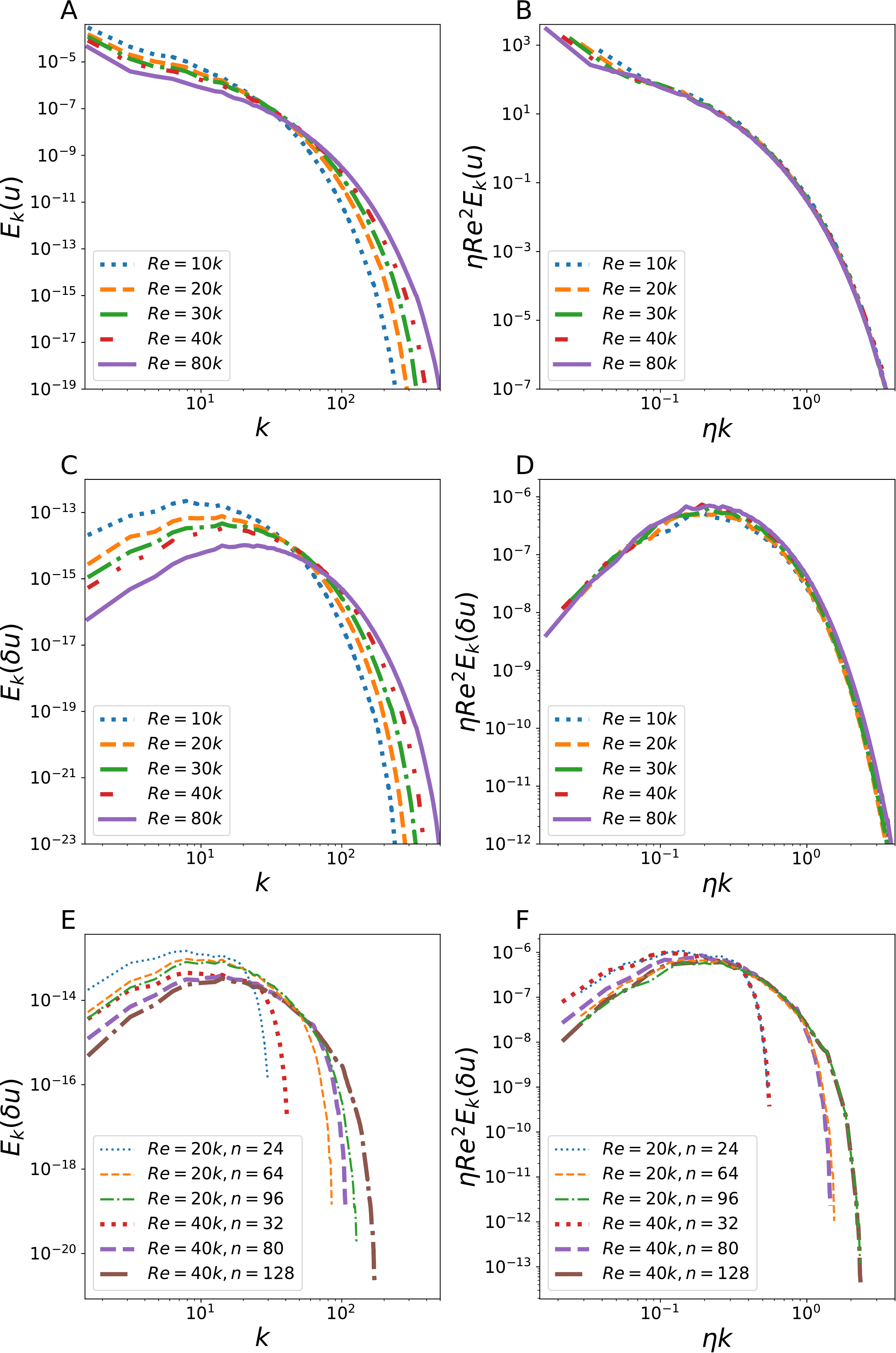}{
    (color online) 
    \textbf{A.}
    Shell-averaged energy spectra of turbulence in DNS. 
    \textbf{B.}
    Same as A, in Kolmogorov units.  
    \textbf{C.}
    Shell-averaged energy spectra of the Lyapunov vectors in DNS. 
    \textbf{D.}
    Same as C, in Kolmogorov units.
    \textbf{E.} Shell-averaged energy spectra of the Lyapunov 
    vectors in LES at $\pRe = 20{,}000$ and $\pRe = 40{,}000$ for a 
    select subset of the cutoff resolutions.
    \textbf{F.} Same as E, in Kolmogorov units. 
}

\figref{spectra}A shows the shell-averaged energy spectra of turbulence
obtained as an ensemble average over $10$ states in each DNS domain 
and the reported uncertainties are one standard deviation.
Following \cite{cerbus2020smallscale}, we replotted these spectra in
\figref{spectra}B in Kolmogorov units to show that they collapse at large
wave numbers (small scales). As we compute the Lyapunov exponent, we expect
the perturbations $\delta \vu$ to align with the most unstable direction of
the system, i.e. the leading Lyapunov vector. Taking these as an
approximation to the leading Lyapunov vector, we plotted their energy
spectra in \figref{spectra}C. Similar to the flow states, when we rescale
these spectra by $(\et \pRe)^2$ and plot against the wave numbers
premultiplied by $\et$, we observe that they similarly collapse onto a
single curve. These suggest that the scaling of the Lyapunov exponents with 
Kolmogorov time $\ta_\et$ can be attributed to the small-scale 
universality of turbulence \cite{schumacher2014smallscale}. 
As shown in \figref{spectra}E and F, we observe a similar collapse of the 
energy spectra for Lyapunov vectors when we compare domains whose resolutions 
match one another in Kolmogorov units. 
We chose the $\pRe$ and resolutions for \figref{spectra}E and F by
noting that $\et|_{\pRe=20{,}000} / \et|_{\pRe=40{,}000} \approx 1.3$
is also approximately equal to the ratio of number of grid points, i.e. 
$32/24 \approx 80/64 \approx 128 / 96 \approx 1.3$. 
In other words, the energy spectra of Lyapunov vectors in two LES simulations 
at different \pRe\ are nearly identical when we use the same LES resolution 
in units of $\et$.

Having obtained the scaling of the LES Lyapunov 
exponents in Kolmogorov units, we replot our data as a function of the gridspacing 
$\Delta$ along with a power law fit in \figref{scale_dep_lyap} 
A and B in linear and log-log scales, respectively.
In \figref{scale_dep_lyap} and the rest of this article, all 
lengths and times are given in Kolmogorov units.
The largest Lyapunpov exponent obtained by LES of different
spatial resolutions $\Delta$ as a function of the latter is
well described by a power law
\eq{powerlawturbulence}{
    \lambda(\Delta) = \al \Delta^{-\rh}
}
where $\al \approx \alnum$ and  $\rh \approx \rhnum$. 
This is the central result: The smaller are the 
spatial scales that the perturbation fields $\delta \vu$ can contain, the 
faster are their growth rates. 
This is also reflected in \figref{spectra}F, where 
it can be seen that increasing the resolution of LES results in 
more and more energy in the
small-scales of the Lyapunov vectors' energy spectra. 
Note, also that as the LES resolution is increased, the 
spectra of the Lyapunov vectors in \figref{spectra}F better approximates 
those obtained in DNS which are shown in \figref{spectra}D.

\fig{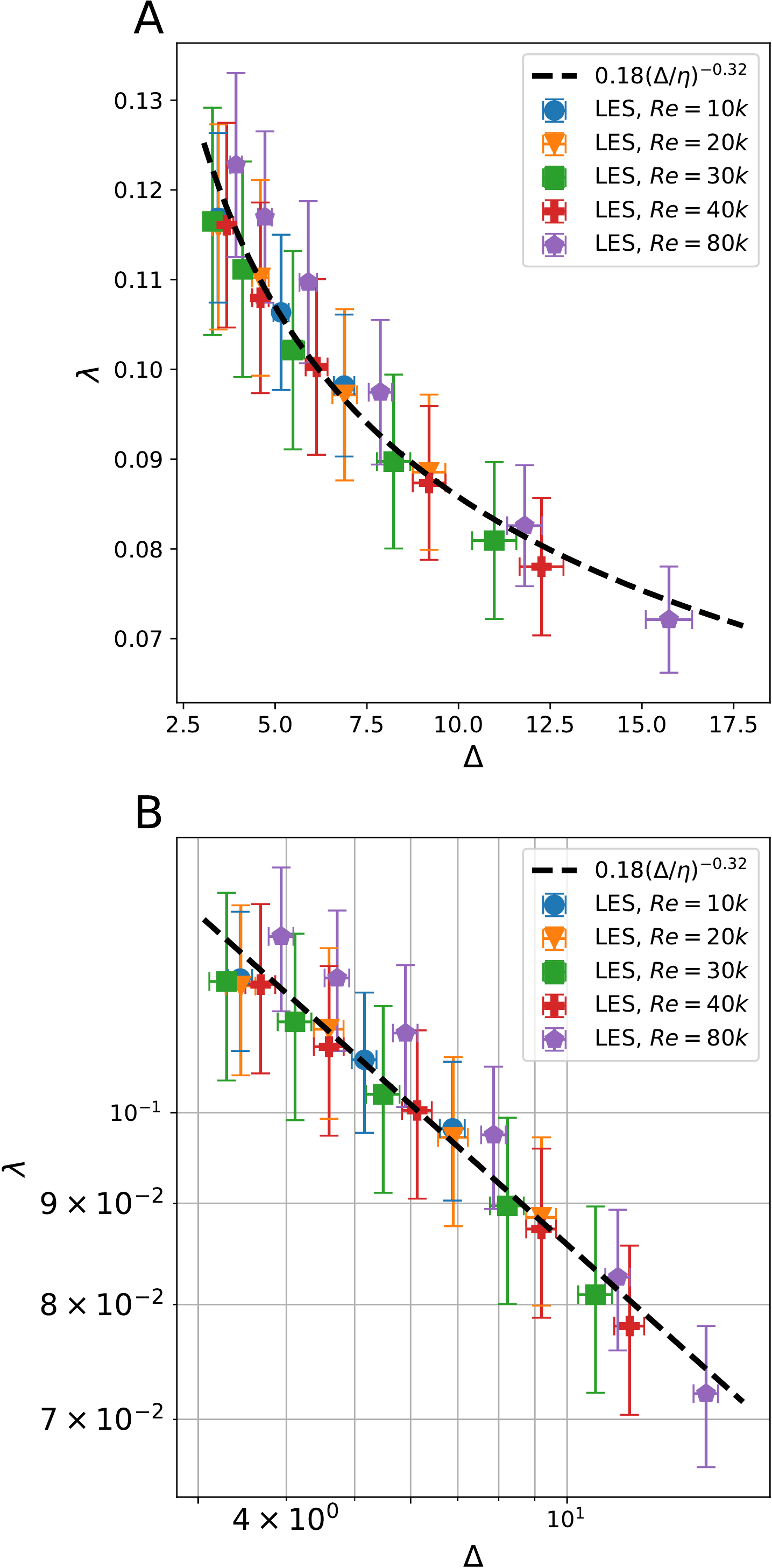}{
    (color online) 
    \textbf{A.} Maximal Lyapunov exponents in LES as a function of grid spacing 
    in Kolmogorov units along with the best-fit power law. 
    \textbf{B.} Log-log plot of the same data. 
}

In the limit $\DE \rightarrow 0$, \eqref{powerlawturbulence} with $\rh > 0$
yields $\la \rightarrow \infty$. Of course, this limit is an unphysical one
since \eqref{powerlawturbulence} is only valid for $\DE \gg \et$. Nevertheless
by examining it, we can derive an upper bound for the prediction horizon that
can be achieved by LES. Let us imagine a forecast
scenario where the initial state of the fluid is determined using a measurement
grid with spacing $\DE$ and modeled with an LES of the same resolution. In such
a setup, it is reasonable to assume that the initial errors of our measurement
will be proportional to $\DE$, i.e. $\ep_0 = \ga \DE$, where $\ga$ accounts for
all the other factors that contribute to the initial error, for instance
temporal resolution, which we assume to remain unchanged. Exponential growth
of this initial error for some time $t$ with the Lyapunov exponent 
\eqref{powerlawturbulence} yields 
\eq{err_exp}{
    \ep (t) = \ga \DE e^{\al \DE^{-\rh} t},
} 
which we can solve for the prediction horizon $t_{pred}$ to reach a maximum tolerable 
$\ep_{th}$ as 
\eq{t_pred}{
    t_{pred} = \frac{\DE^{\rh}}{\al}  
        \left[\ln \ep_{th} - \ln (\ga \DE) \right] .
}
This is the standard expression for the prediction horizon with 
$\DE^{\rh} / \al$ and $\ga \DE$ taking the role of the Lyapunov time and 
the initial error, respectively. Note that when $\DE \rightarrow 0$ we get 
$t_{pred} = 0$ due to the vanishing of the Lyapunov time, which already 
tells us decreasing the grid spacing $\DE$ might not necessarily translate 
to an extended prediction horizon. Under which condition this is the case 
is given by the derivative 
\eq{ddDE_t_pred}{
    \frac{d t_{pred}}{d \DE}  = 
        \frac{\DE^{\rh - 1}}{\al}  
        \left[ 
            \rh \ln \frac{\ep_{th}}{\ga \DE} - 1 
        \right]\,, 
}
which is negative when $\ep_{th} / (\ga \DE) < e^{1/\rh}$ and the equality 
gives the point of diminishing returns because if 
$ d t_{pred} / d \DE > 0$, then reducing $\DE$ results in a shorter 
prediction horizon. Consequently, the optimal resolution is given by 
the $0$ of \eqref{ddDE_t_pred} as 
\eq{opt_res}{
    \DE^* = \frac{\ep_{th}}{e^{1/\rh} \ga} 
}
Plugging it into \eqref{t_pred}, we get the maximal prediction horizon
\eq{max_t_pred}{
    t^*_{pred} = \frac{\ep_{th}^{\rh}}{e \al \rh \ga^{\rh}} .
}
We illustrate \eqref{t_pred} in \figref{t_pred} where we 
plotted it as a function of $\DE$ for $\ep_{th} = 1000$ (arbitrary) and 
$\gamma = 1,2,3,4$ along with $t^*$ indicating the maximal prediction 
horizon as a function of $\ga$.
In the light of our results, $t^*$ given by \eqref{max_t_pred} is the 
longest prediction horizon that is achievable through LES, if the
magnitude of initial uncertainties is proportional to the LES grid
spacing $\DE$ and $\DE \gg \eta$, i.e. the LES grid is much coarser
than the Kolmogorov length. This can be straightforwardly
generalized to initial errors that scale differently with the grid
spacing, i.e. $\ep_0 = \ga \DE^\mu$, with $\mu > 0$.  
Following steps analogous to (\ref{err_exp}--\ref{ddDE_t_pred}), we get
the derivative of prediction horizon with respect to the grid spacing 
for this case as
\eq{ddDE_t_pred_mu}{
    \frac{d t_{pred} (\mu)}{d \DE}  = 
        \frac{\DE^{\rh - 1}}{\al}  
        \left[ 
            \rh \ln \frac{\ep_{th}}{\ga \DE} - \mu 
        \right]\,, 
}
which is negative as long as $\ep_{th} / (\ga \DE) < e^{\mu/\rh}$. 
Thus, in principle, the prediction horizon can be extended by ensuring 
this condition if $\mu > 1$ is technologically possible. Conversely, if 
reducing the grid spacing results in a slower-than-linear improvement 
($\mu < 1$) of the initial condition's precision, then the point of 
diminishing returns is reached earlier. 

\fig{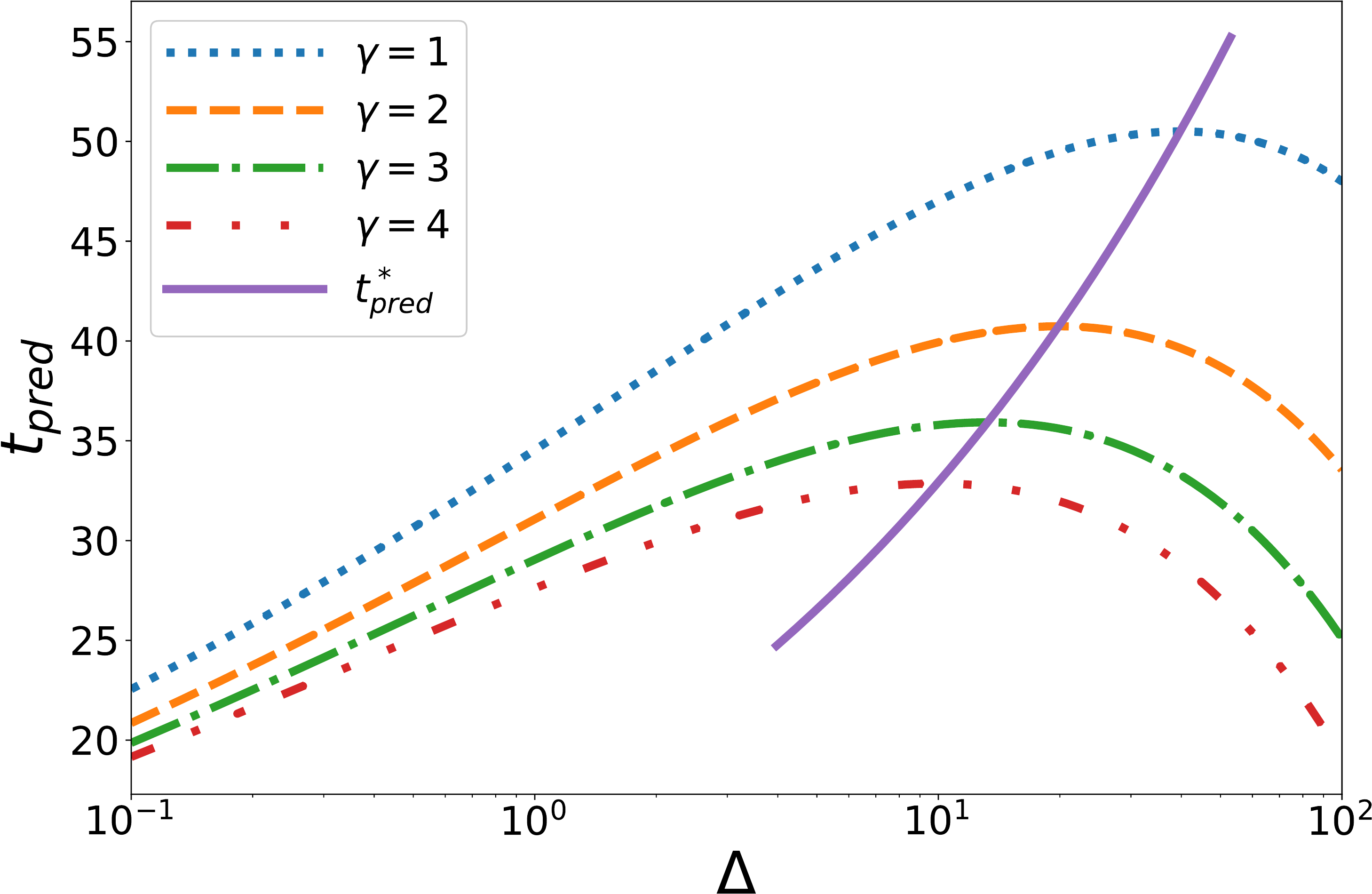}{
    (color online) Prediction horizon \eqref{t_pred} as a function of grid  
    spacing plotted for $\gamma = \{1, 2, 3, 4\} \mbox{ and } \ep_{th} = 1000$.
}

\section{Conclusion}
    
In view of weather forecasting, this might have severe implications for the
limits of predictions. Numerical weather prediction models and atmospheric
general circulation models (AGCMs) contain the  Navier--Stokes equations in some
form as the core for modeling atmospheric transport
\cite{coiffier2011fundamentals,warner2017numerical} among the so-called
primitive equations. For better resolution of small scale atmospheric processes
with the ultimate goal to model convection and clouds, the grid spacing in
weather models has been reduced according to compute power availability and is
currently down to, e.g., 2.2km in the high-resolution DWD model for
Germany\cite{DWD-ICON-D2}. It is well known by practitioners that high-resolution models
lose their predictability on their small scales much faster than coarser
models do on their respective scales, and consequently high resolution
models are used for short term forecasts only (e.g., DWD limits the
forecast runs of ICON-D2 to 27 hours). Our finding of an optimal
resolution for maximising the prediction horizon sets this empirics
into a broader context and shows that forecast horizon cannot be
extended by simply improving resolution alone.

Scaling of the maximal Lyapunov exponents of a turbulent flow with the 
Kolmogorov time scale was conjectured in \cite{ruelle1995microscopic}, 
assumed in the subsequent literature \cite{aurell1996growth} and recently 
contested in homogeneous isotropic turbulence simulations 
\cite{boffetta2017chaos,mohan2017scaling,berera2018chaotic}. 
While our results are in agreement with the 
conjecture, thus at odds with the results of 
\cite{boffetta2017chaos,mohan2017scaling,berera2018chaotic}, 
we cannot rule out subtle variations or finite-\pRe\ effects on this scaling  
based on the present results. 
We would like to note, however, two important differences between 
the cited studies and the present one. Firstly, the resolution of simulations 
we perform are roughly twice the ones in 
\cite{boffetta2017chaos,mohan2017scaling,berera2018chaotic}. We chose to perform 
DNS at resolutions much higher than the traditional turbulence literature following the 
findings of \cite{donzis2008dissipation,schumacher2014smallscale} which emphasized 
importance of resolving the Kolmogorov scale for observing the small-scale universality 
in simulations. The other potential explanation of the apparent discrepancy is  
    \cite{boffetta2017chaos,mohan2017scaling,berera2018chaotic}'s use of 
    Machiel's \cite{machiels1997predictability} forcing which is a positive feedback 
    on the large scales. Although irrelevant for small scales, such a forcing term 
    could make the large scales artificially unstable and result in an 
    unexpected behavior of the Lyapunov exponents. 
Another recent paper \cite{nastac2017lyapunov}
computing Lyapunov exponents in LES and DNS of homogeneous isotropic turbulence 
reported a DNS estimate $\la \ta_\et \approx 0.122$, which is within the 
error bars of our results in \figref{lyap_les_dns}, suggesting a universal 
behavior of the maximal Lyapunov exponent. 
Altogether, we believe that further research is necessary for settling the 
question whether the maximal Lyapunov exponent in turbulence indeed scales 
with the Kolmogorov time. 

In summary, we studied the rate of error growth in LES and DNS of sinusoidally
forced Navier--Stokes equations in three-dimensions at different Reynolds
numbers and LES resolutions. We found that independent of the Reynolds number in
the turbulent regime, the largest Lyapunov exponent is a fixed multiple of the
inverse Kolmogorov time in DNS, and the LES exponents 
at different $\pRe$ collapse onto a single curve that can be approximated 
by a power law when nondimensionalized using Kolmogorov units. 
Using this power law as a phenomenological model of the scale-dependent 
error growth, we showed that in a forecast scenario where initial errors are 
proportional to the resolution cutoff, the scale dependent error growth introduces 
an upper limit to the achievable prediction horizon.
While the exact functional dependence and the numerical values 
of LES Lyapunov exponents are likely to
depend on the studied models, we
expect the scale-dependent error growth to be a common feature of turbulence
since it appears to be related to the small scales, at which turbulent flows
exhibit universality \cite{schumacher2014smallscale}. We thus believe that
scale-dependent error growth should be taken into consideration in forecast
scenarios where hydrodynamic transport is modelled. 

\textit{Acknowledgments.}
We gratefully acknowledge the computing resources provided by the 
Max Planck Computing \& Data Facility in Garching, Germany.

\bibliography{lyascale}

\end{document}